\documentclass[12pt]{article}
\usepackage[latin1]{inputenc}
\usepackage[english]{babel}
\usepackage{textcomp}
\usepackage{bbm}
\usepackage[dvips]{geometry}
\usepackage{graphicx}
\usepackage[dvips]{color}
\usepackage[section]{placeins}
\usepackage[ruled,vlined]{algorithm2e}
\usepackage{subfigure}%Diverses figures a sa mateixa
\usepackage[sectionbib]{natbib}
\bibpunct{(}{)}{;}{a}{,}{,}
\begin{document}
% Le titre du papier

\title{
Parallel machine scheduling with precedence constraints and setup times
}
% Mettre ci-dessous le titre qui sera affiche en entte des pages impaires: il s'agit du titre 
% ou d'une version abrge
%\titlerunning{Tree and local search for sheduling problems}
% Mettre ci-dessous le titre qui sera affiche en entte des pages impaires: il s'agit du titre 
% ou d'une version abrge
\author{Bernat Gacias $^{1,2}$, Christian Artigues $^{1,2}$ and Pierre Lopez $^{1,2}$\\
~\\
\begin{footnotesize}$^{1}$~CNRS; LAAS; 7 avenue du Colonel Roche, F-31077 Toulouse, France\end{footnotesize}\\ 
\begin{footnotesize}$^{2}$~Université de Toulouse; UPS, INSA, INP, ISAE; LAAS; F-31077 Toulouse, France\end{footnotesize} \\
\begin{small}\{bgacias,artigues,lopez\}@laas.fr\end{small}}
\date{}
\maketitle
\begin{abstract}
This paper presents different methods for solving parallel machine scheduling problems with precedence constraints and setup times between the jobs. Limited  discrepancy search methods mixed with local search principles, dominance conditions and specific lower bounds are proposed. The proposed methods are evaluated on a set of randomly generated instances and compared with previous results from the literature and those obtained with an efficient commercial solver. We conclude that our propositions are quite competitive and our results even outperform other approaches in most cases. 
\end{abstract}
~\\
%\small 
\begin{footnotesize}\textbf{Keywords}: Parallel machine scheduling, setup times, precedence constraints, limited discrepancy search, local search.\end{footnotesize}
\thispagestyle{empty}
\section{Introduction}
%~\\[-5mm]
~~~This paper deals with parallel machine scheduling with precedence constraints and setup times between the execution of jobs. We consider the optimization of two different criteria: the minimization of the sum of completion times and the minimization of maximum lateness. These two criteria are of great interest in production scheduling. The sum of completion times is a criterion that maximizes the production flow and minimizes the work-in-process inventories. Due dates of jobs can be associated to the delivery dates of products. Therefore, the minimization of maximum lateness is a goal of due date satisfaction in order to disturb as less as possible the customer who is delivered with the longest delay. These problems are strongly \emph{NP-hard}~\citep{bib-NP}.

The parallel machine scheduling problem has been widely studied~\citep{bib-parallel}, specially because it appears as a relaxation of more complex problems like the hybrid flow shop scheduling problem or the RCPSP (Resource-Constrained Project Scheduling Problem). Several methods have been proposed to solve this problem. In~\cite{bib-GenColonnes}, a column generation strategy is proposed. \cite{bib-LinearProgram} propose a linear program and an efficient heuristic for large-size instances for the resolution of priority constraints and family setup times problem.~\cite{bib-lbCtotal} solve the problem with a tree search method. More recently,~\cite{bib-TreeSearch} compare two different branching schemes and several tree search strategies for the problem with release dates and tails for the makespan minimization case. 

However, the literature on parallel machine scheduling with precedence constraints and setup times is quite limited.~\cite{bib-PrecSommeCi} and~\cite{bib-PrecLmax} deal with the problem with precedence constraints for the minimization of the sum of completion times and maximum lateness respectively. The setup times case is considered in~\cite{bib-SetupLmax} and in ~\cite{bib-SetupLmax2} for the minimization of maximum lateness.~\cite{bib-Lmax} deal with the same criterion on a single machine with family-dependent setup times. Finally,~\cite{bib-lbCi} propose a lower bound and a branch-and-bound method for the minimization of the sum of completion times. 

Problems that have either precedence constraints or setup times, but not both, can be solved by list scheduling algorithms. It means there exists a total ordering of the jobs (i.e., a list) that, when a given machine assigment rule is applied, reaches the optimal solution~\citep{bib-ListScheduling}. For a regular criterion, this rule is called Earliest Completion Time (ECT). It consists in allocating every job to the machine that allows it to be completed at the earliest. This reasoning unfortunately does not work when precedence constraints and setup times are considered together, as shown in~\cite{bib-contraexemple}. We have then to modify the way to solve the problem and consider both scheduling and resource allocation decisions.

In Section 2, we define formally, the parallel machine scheduling problem with setup times and precedence constraints between jobs. In Section 3 we present a branch-and-bound method and its components: tree structure, lower bounds, and dominance rules. Discrepancy-based tree search methods are described in Section 4. In Section 5 we present the hybrid tree-local search methods used to solve large-size instances. Section 6 is dedicated to computational experiments.
\section{Problem definition}\label{ProblemDefinition}
~~~We consider a set $J$ of $n$ jobs to be processed on $m$ parallel machines. The precedence relations between the jobs and the setup times, considered when different jobs are sequenced on the same machine, must be satisfied. The preemption is not allowed, so each job is continually processed during $p_{i}$ time units on the same machine. The machine can process no more than one job at a time. The decision variables of the problem are the start times of every job $i=1..n$, $S_{i}$, and let us define $C_{i}$ as the completion time of job $i$, where $C_{i}=S_{i}+p_{i}$. Let $r_{i}$ and $d_{i}$ be the release date and the due date of job $i$, respectively. Due dates are only considered for job lateness computation. We denote by $E$ the set of precedence constraints between jobs. The relation $(i,j)\in E$, with $i, j \in J$, means that job $i$ is performed before job $j$ ($i \prec j$) such that job $j$ can start only after the end of job $i$ $(S_{j}\geq C_{i})$. Finally, we define $s_{ij}$ as the setup time needed when job $j$ is processed immediately after job $i$ on the same machine. Thus, for two jobs $i$ and $j$ processed successively on the same machine, we have either $S_{j}\geq C_{i}+s_{ij}$ if $i$ precedes $j$, or $S_{i}\geq C_{j}+s_{ji}$ if $j$ precedes $i$. Using the notation of~\cite{bib-NP}, the problems under consideration are denoted: $Pm|prec,s_{ij},r_i|\sum C_{i}$ for the minimization of the sum of completion times and $Pm|prec,s_{ij},r_i|L_{\max}$ for the minimization of the maximum lateness.
\subsection*{Example}
\begin {figure}[h]
\begin {center}
\scalebox{0.4}{
\includegraphics[]{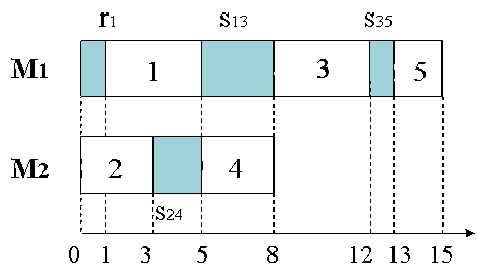}}
\caption{Feasible schedule}
\label{fig-exemple}
\end {center}
\end {figure}

~~~A set of $5$ jobs $(n=5)$ must be executed on $2$ parallel machines $(m=2)$. For every job $i$, we give $p_{i}$, $r_{i}$, $d_{i}$, and $s_{ij}$ (see Table~\ref{tab-donnes}). Besides, for that example we have the precedence constraints:  $1\prec 4$ and $2\prec 5$.
%\vspace{-0.1 cm}  
\begin {table}[h]
\begin{center}
	\subtable[]{
	\begin {tabular}{cccc}
	\hline
	\bf{$n$} & \bf{$p_{i}$} & \bf{$r_{i}$} & \bf{$d_{i}$}\\
	\hline
	\bf{1}	&		4 	  & 	1 		&	  7\\
	%\hline
	\bf{2}	&		3	  &	0		&	  5\\	
	%\hline
	\bf{3}	&		4	  &	3		&    8\\
	%\hline
	\bf{4}	&		3	  &	3		&	 10\\
	%\hline
	\bf{5}	&		2	  &	1		&	  5\\
	\hline
	\end {tabular}}
	\hspace{1.5 cm}
	\subtable[]{
	\begin {tabular}{cccccc}
	\hline
	\bf{$s_{ij}$}&\bf{1}&\bf{2}&\bf{3}&\bf{4}&\bf{5}\\
	\hline
	\bf{1}	&		0 	  & 	2 		&	  3	&	4	&	5\\
	%\hline
	\bf{2}	&		7	  &	0		&	  6	&	1	&	3\\	
	%\hline
	\bf{3}	&		2	  &	4		&    0	&	7	&	1\\
	%\hline
	\bf{4}	&		4	  &	4		&	 8	&	0	&	1\\
	%\hline
	\bf{5}	&		3	  &	4		&	 8	&	5	&	0\\
	\hline
	\end {tabular}}
\caption {Example 1 data}
\label {tab-donnes}
\end {center}
\end {table}
\FloatBarrier
Figure~\ref{fig-exemple} displays a feasible solution for this problem. The set of precedence constraints is satisfied: $S_{5}=13\geq 3=C_{2}$ and $S_{4}=5\geq 5=C_{1}$. We stress that job $4$ must postpone its start time on $M_{2}$ by one time unit because of the precedence constraint. On the other hand, we have to check that, for every job $i$, $r_{i}\leq S_{i}$ and that setup times between two sequenced jobs on the same machine are also respected. For the evaluation of the solution, we observe that for the minimization of the sum of completion times the value of the function is $z=\sum C_{i}=43$ and for the minimization of maximum lateness $z=L_{\max}=L_{5}=10$.
\FloatBarrier
\section{Branch-and-Bound components for  
%\linebreak
$Pm|prec,s_{ij},r_i|\sum{C_i}$ and $Pm|prec,s_{ij},r_i|L_{\max}$}\label{sec-branchandbound}
~~~~A tree structure with two levels of decisions (scheduling and resource allocation) is defined in Section~\ref{subsec-structure}. Lower bounds, constraint propagation mechanisms and dominance rules are introduced in Sections~\ref{subsec-evaluation} and~\ref{subsec-dominance}.
\subsection{Tree structure}\label{subsec-structure}
\begin{figure}
\begin{center}
\subfigure[Optimal schedule without the precedence constraint]{
\scalebox{0.4}{
\includegraphics[]{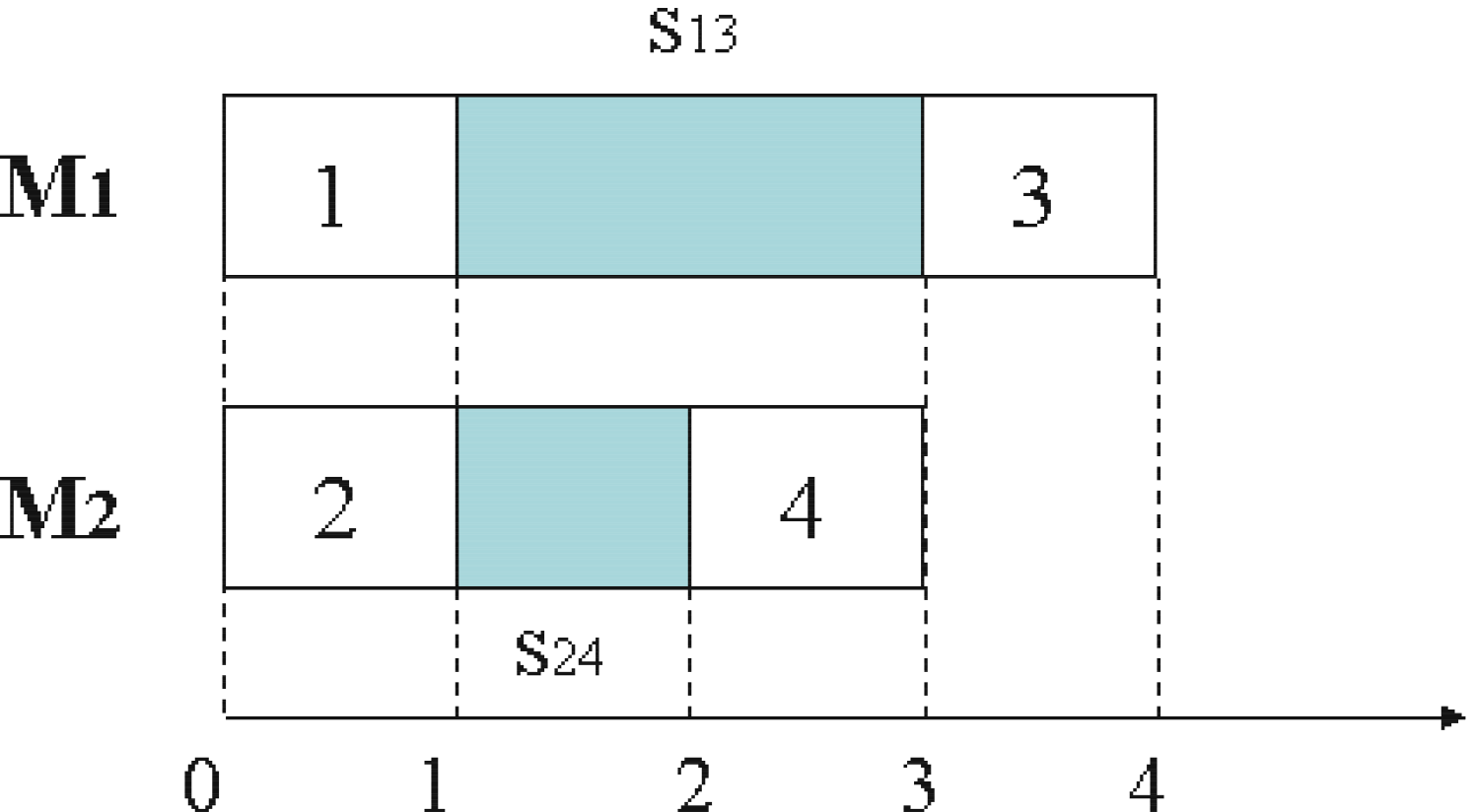}
}
}
\subfigure[Optimal schedule with the precedence constraint]{
\scalebox{0.4}{
\includegraphics[]{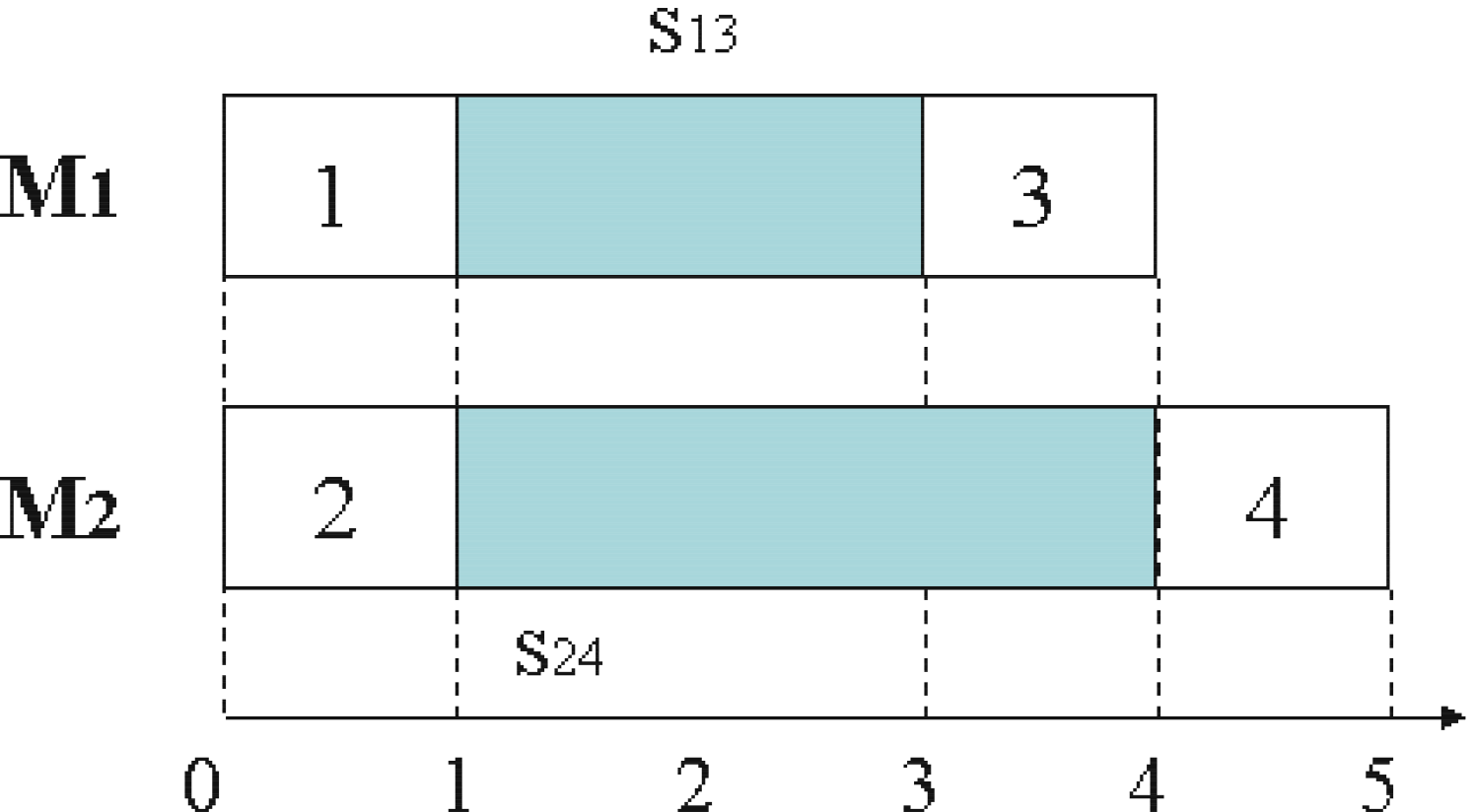}
}
}
\caption{Example of job allocation}
\label{fig-contraexemple}
\end{center}
\end {figure}

~~~~Precedence constraints and setup times scheduling problems may not be efficiently solved by a list algorithm as conjectured by~\cite{bib-contraexemple}. It means that there possibly does not exist a job allocation rule that reaches an optimal solution where all the possible lists of jobs are enumerated. Let us consider the minimization of the sum of completion times for 4 jobs scheduled on 2 parallel machines. The data of the problem are displayed in Table 2.
\begin {table}[h]
\begin{center}
	\subtable[]{
	\begin {tabular}{cccc}
	\hline
	\bf{$n$} & \bf{$p_{i}$} & \bf{$r_{i}$}\\
	\hline
	\bf{1}	&		1 	& 0\\
	%\hline
	\bf{2}	&		1	  &	0\\	
	%\hline
	\bf{3}	&		1	  &	2\\
	%\hline
	\bf{4}	&		1	  &	2\\
	\hline
	\end {tabular}}
	\hspace{1.5 cm}
	\subtable[]{
	\begin {tabular}{cccccc}
	\hline
	\bf{$s_{ij}$}&\bf{1}&\bf{2}&\bf{3}&\bf{4}\\
	\hline
	\bf{1}	&		0 	& 10 	&	  2	&	10\\
	%\hline
	\bf{2}	&		10	  &	0		&	  1	&	1\\	
	%\hline
	\bf{3}	&		10	  &	10		& 0	&	10\\
	%\hline
	\bf{4}	&		10	  &	10		&	 10	&	0\\
	\hline
	\end {tabular}}
\caption {Example 2 data}
\label {tab-donnes2}
\end {center}
\end {table}
\FloatBarrier
If we consider the problem without precedence constraints, we find two optimal solutions ($\sum C_i=9$) when we allocate the jobs following the Earliest Completion Time rule for the lists $\{1,2,4,3\}$ and $\{2,1,4,3\}$ (see Figure~\ref{fig-contraexemple}a). Now, let us consider the same problem with the precedence constraint $3\prec4$. In that case, there does not exist any allocation rule that reaches an optimal solution for any list of jobs that respects the precedence constraint. The optimal solution ($\sum C_i=11$) is reached when we consider the list $\{1,2,3,4\}$ and job 3 is not allocated on the machine that allows it to finish first (see Figure~\ref{fig-contraexemple}b). Thus, in our problems we have not only to find the best list of jobs but also to specify the best resource allocation.    
\FloatBarrier
The optimal solution can be reached by a two decision-level tree search. We define a node as a partial schedule $\sigma(p)$ of $p$ jobs. Every node entails at most $m\times(n-p)$ child nodes. The term $n-p$ corresponds to the choice of the next job to be scheduled (job scheduling problem). Only the jobs with all the previous jobs already executed are candidates to be scheduled. Once the next job to be scheduled is selected we have to consider the $m$ possible machine allocations (machine allocation problem). For practical purposes, we have mixed both levels of decision: one branch is associated with the choice of the next job to schedule and also with the choice of the machine. A solution is reached when the node represents a complete schedule, that means when $p=n$.
%We suggest the following proposition in order to reduce the number of nodes to explore: for every job $x$ having $t$ (direct or %undirect) successor jobs, we consider the assignments on the first $[\min(m,t+1)]^{th}$ machines that allows $x$ to be completed %as soon as possible. So, we are going to consider the schedule according to \emph{ECT} rule for all jobs, except for the previous %jobs. In that case, we consider to schedule them on more than one machine to prevent that the previous jobs could avoid the best %assignment for its successor jobs.
\subsection{Node evaluation}\label{subsec-evaluation}
~~~~Node evaluation differs depending on the studied criterion. First, we propose to compute a simple lower bound. For every node (partial schedule), we update the earliest start times of the unscheduled jobs taking account of the branching decisions through precedence constraints and we calculate the minimum completion time (for $\min \sum C_{i}$ criterion) and the minimum lateness (for $\min L_{\max}$ criterion) for every not yet-scheduled job. Then we update the criterion and we compare the lower bound with the best current solution.

We propose to compute an upper bound. The upper bound is computed by a simple list scheduling heuristic selecting the combination of job, between the not yet-scheduled jobs, and machine with the shortest start time.  

For criterion $\min \sum C_{i}$, we also propose to compute the lower bound presented in~\cite{bib-lbCi} for the parallel machine scheduling problem, with sequence-dependent setup times and release dates ($Pm|s_{ij},r_i|\sum C_{i}$). This problem is a relaxation of the problem with precedence constraints, so the lower bound is still valid for our problem. In this paper, we just present the lower bound for the problem, that is based on job preemption relaxation, and we refer to~\cite{bib-lbCi} for the proof. 

Let $S_*$ be the schedule obtained with the SRPT (Shortest Remaining Processing Time) rule for the relaxed problem $1|r_i,(\frac{p_i}{m}+s^*_i),pmtn|\sum{\max(C^*_i-s^*_i,r_i+p_i)}$, where $s_i=\min_{j\neq i} s_{ij}$ and $s^*_i=\frac{s_i}{m}$. Let $C^*_{[i]}(S_{*})$ be the modified completion time of job $i$ with the processing time $p_i+s^*_i$ for each job $i$. Let $a_i=p_i+r_i+s^*_i$ and let ($a_{[1]},a_{[2]},\dots,a_{[n]}$) be the series obtained by sorting ($a_1,a_2,\dots,a_n$) in non-decreasing order. Then $LB=\sum{\max[C^*_{[i]}(S_{*}),a_{[i]}]}-\sum{s^*_i}$ is a lower bound for $Pm|prec,s_{ij},r_i|\sum C_{i}$. The complexity of the lower bound is $O(n\log n)$, the same complexity as SRPT.

For $\min L_{\max}$, the evaluation consists in triggering a satisfiability test based on constraint propagation involving energetic reasoning~\citep{bib-energetique2}. The energy is produced by the resources and it is consumed by the jobs. We apply this feasibility test to verify whether the best solution reached from the current node will be at least as good as the best current solution. We determine the minimum energy consumed by the jobs ($E_{consumed}$) over a time interval $\Delta=[t_{1},t_{2}]$ and we compare it with the available energy ($E_{produced}=m\times(t_{2}-t_{1})$). In our problem we also have to consider the energy consumed by the setup times ($E_{setup}$). If $E_{consumed}+E_{setup}>E_{produced}$ we can prune the node.

For an interval $\Delta$ where there is a set $F$ of $k$ jobs that may consume energy, we can easily show that the minimum quantity of setups which occurs is $\alpha=\max(0,k-m)$. So, we have to take the $\alpha$ shortest setup times of the set ${\{s_{ij}\},i,j\in F}$, into account. 

The energy consumed in an interval $\Delta$ is $E_{consumed}=\sum_{i}\max(0,\min(p_{i},t_{2}-t_{1},r'_{i}+p_{i}-t_{1},t_{2}-d'_{i}+p_{i}))+\sum_{l}^{\alpha}s_{[l]}$ where $s_{[l]}$ are the setup times of the set ${\{s_{ij}\},i,j\in F}$, sorted in non-decreasing order, and a time window $[r'_{i},d'_{i}]$ for every not yet-scheduled job $i$ is issued from precedence constraint propagation: 
\begin{center}
$r'_{i}=\max \{r_{i},r_j+p_j;~\forall~j\in\Gamma_{i}^{-}\}$ and $d'_{i}=\min \{Z_{best}+d_{i},d'_j-p_j;~\forall~j\in\Gamma_{i}^{+}\}$, 
\end{center}
where $\Gamma_{i}^{-}$ and $\Gamma_{i}^{+}$ are respectively the set of previous and successor jobs for job $i$ and $Z_{best}$ is the minimum current value for $L_{\max}$.

\begin {figure}[h]
\begin {center}
\scalebox{0.3}{
\includegraphics[]{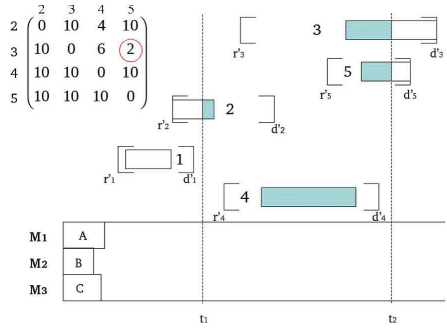}
}
\caption{Minimum energy consumed in a partial schedule}
\label{fig-energie}
\end {center}
\end {figure}

In Figure~\ref{fig-energie} we illustrate how to compute the energy consumed by the not yet-scheduled jobs (1 to 5 in the example) for a 3-machine problem. For every job, we determine a time window and the minimum energy consumed (in grey) over the selected interval $\Delta=[t_{1},t_{2}]$. For $E_{setup}$ we have to take the $\alpha$ shortest setup times, in the example $k=4$ (there is no consumption for job 1) and $m=3$, so we have to sum only the shortest setup time between the consuming jobs, in our case we add 2 energy units (value of $s_{35}$).   
\FloatBarrier
The time interval $\Delta=[t_{1},t_{2}]$ considered to compute the energy consumed is $t_{1}=\min{r'_i},\forall i \in F$ and $t_{2}=d'_j$, where $j$ is the job with the shortest time window $\min{(d'_j-r'_j)},\forall j \in F$. The complexity of the energetic test is $O(n^2)$.  

\subsection{Dominance rules}\label{subsec-dominance} 
~~~We also propose dominance rules to restrict the search space. They consist in trying to find whether there exists a dominant node allowing us to prune the evaluated node. All proposed rules are based on the dominance properties of the set of active schedules. A schedule $S$ is active if no feasible schedule can be obtained from $S$ by left-shifting a single activity. Let us define the \emph{front} of a partial schedule as the set of the last jobs executed on the machines (the ones with the largest start times).    

\begin {figure}[h]
\begin {center}
\scalebox{0.4}{
\includegraphics[]{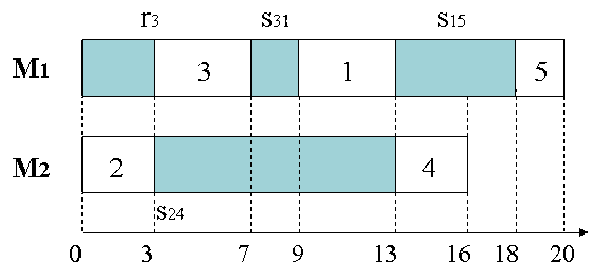}}
\caption{Partial schedule of the evaluated node}
\label{fig-ExempleFlot}
\end {center}
\end {figure}

We first present a global dominance rule based on max flow computation based on a resource-flow model previously used for the resource-constrained project scheduling problem with setup times~(\cite{bib-flotmax2}, Section 2.13). The idea is to verify that there exists a partial schedule $\sigma'(p)$ with the start times \linebreak $S'=\{S'_{1},S'_{2},\dots,S'_{i},\dots,S'_{p}\}$ different from $\sigma(p)$ with start times \linebreak $S=\{S_{1},S_{2},\dots,S_{i},\dots,S_{p}\}$ that allows us to move forward the start time of job $k$ without modifying other start times ($S'_{i}=S_{i},\forall i\neq k$ and $S'_{k}\leq S_{k}-1$). This is a necessary but not a sufficient condition for the dominance. Besides, the schedule $\sigma'(p)$ has to keep the same front as $\sigma(p)$ except for the case where job $k$ does not belong to the front of $\sigma'(p)$ (the dominant partial schedule). For example in Figure~\ref{fig-ExempleFlot}, job 5 ($S_5=18$) may be scheduled after job 4 or between job 2 and job 4 with a shortest start time ($S'_5=17$). In the first case the new schedule $\sigma'(p)$ is not dominant because of setup times but in the second case it is, so the front can be modified only if job $k$ is not part of it in $\sigma'(p)$. 

\begin {figure}[h]
\begin {center}
\scalebox{0.45}{
\includegraphics[]{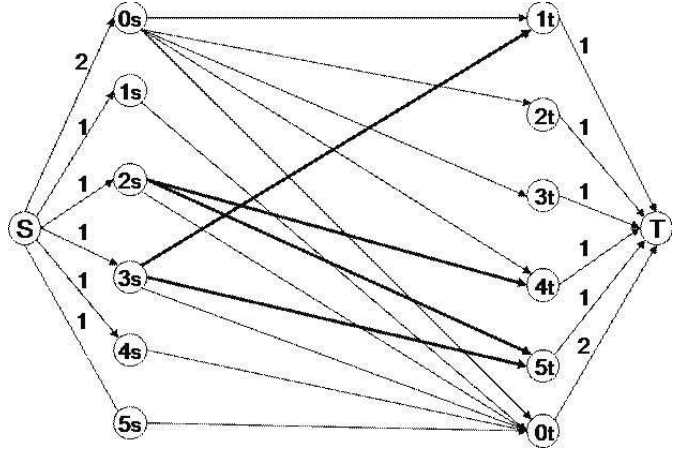}
}
\caption{Network to compute the max flow dominance rule}
\label{fig-flotmax}
\end {center}
\end {figure}

\begin{figure}[h]
\begin{center}
\subfigure[Evaluated node]{
\scalebox{0.3}{
\includegraphics[]{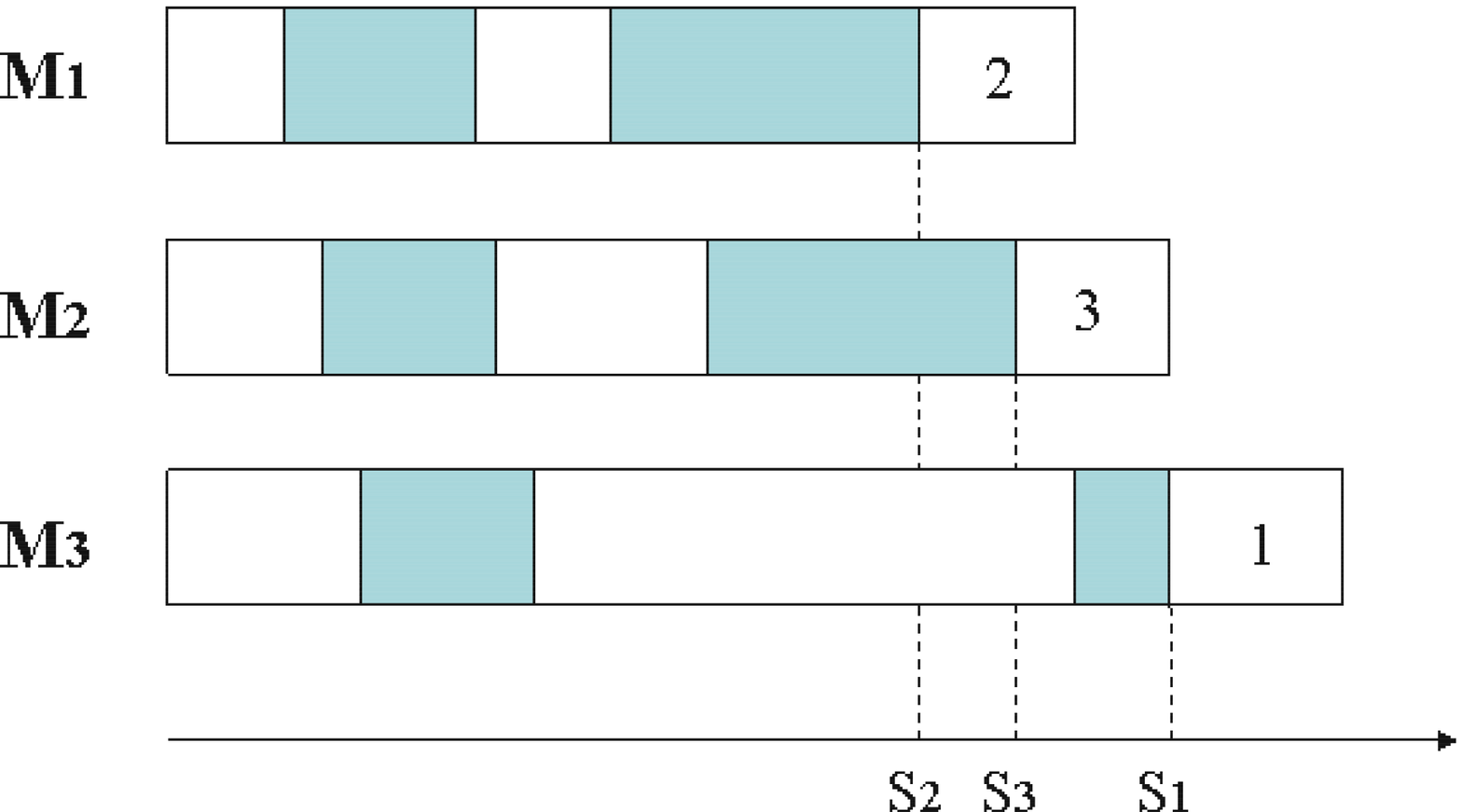}
}
}
\subfigure[Dominant partial schedule]{
\scalebox{0.3}{
\includegraphics[]{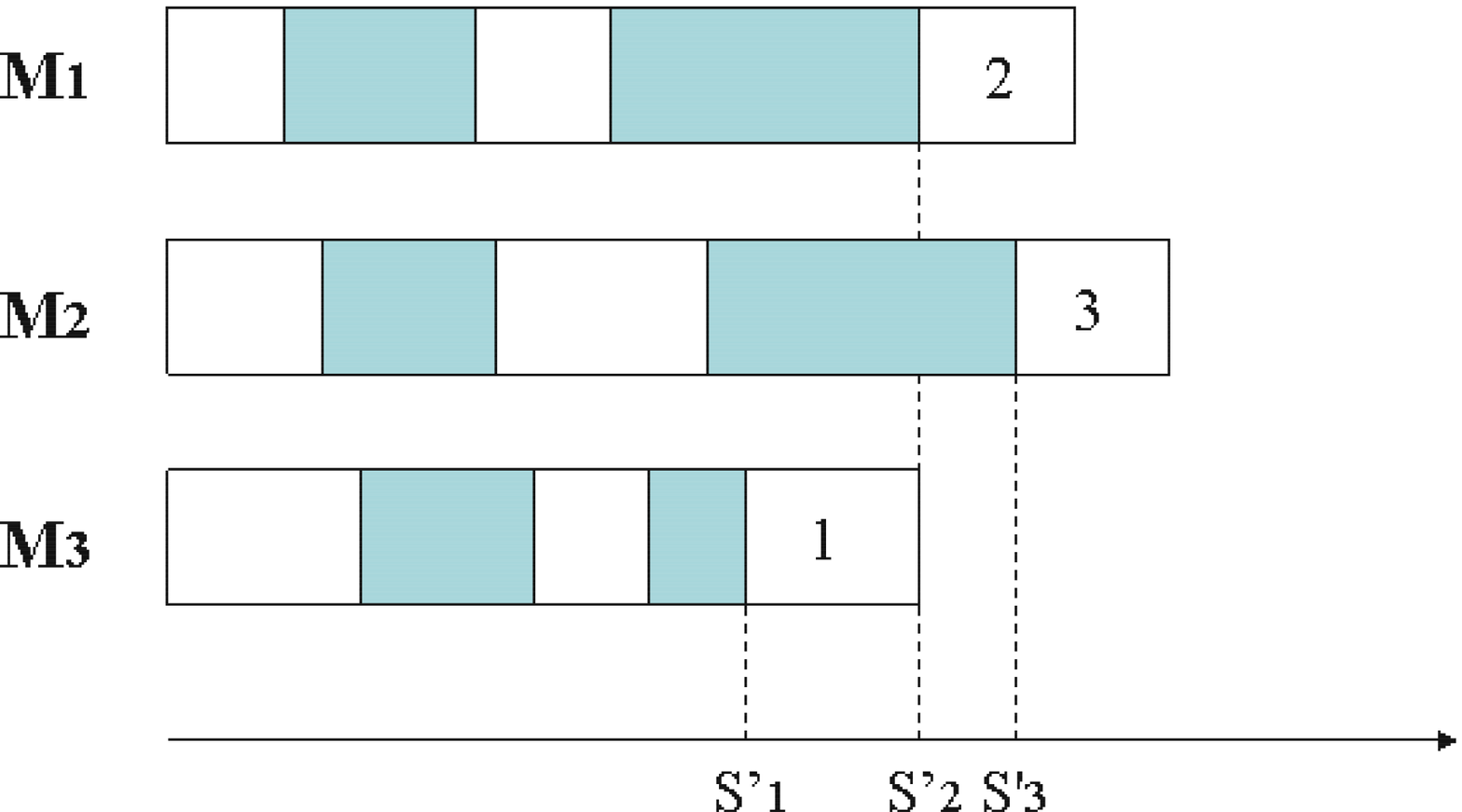}
}
}
\caption{Example of dominant partial schedule}
\label{fig-dominance}
\end{center}
\end {figure}

We represent $\sigma'(p)$ by a graph and we turn the dominance rule in a max flow computation. Two vertices are considered for every job, the first one represents the start time $i_{t}$ and the second one the completion time $i_{s}$ of the job. One unit capacity arcs are defined between the vertices $i_{s}$-$j_{t}$ by the partial schedule $\sigma'(p)$ and they represent the transfer of resource units between the jobs. Finally, we need four dummy vertices. Two vertices ($0_{s}$, $0_{t}$), the source node $S$ and the sink node $T$, flow origin and flow destination, respectively. Arcs $S$-$0_{s}$ and $0_{t}$-$T$ have $m$-unit capacity and represent the resource constraint. 1-unit capacity arcs between $S$-$i_{s}$ and $i_{t}$-$T$ ensure the job execution.
\FloatBarrier  
Figure~\ref{fig-flotmax} shows the flow network for the schedule depicted in Figure~\ref{fig-ExempleFlot} (data of Table~\ref{tab-donnes}). For each node we try to find a schedule that allows us to move forward the start time of the last scheduled job by one unit (job 5 in the example, $S'_{5}=17$) and to keep the same start times for the other jobs. We create a direct arc $i_{s}$-$j_{t}$ if $S'_{j}>S'_{i}+p_{i}+s_{ij}$, that means if job $j$ can be executed on the same machine than job $i$. In order to respect the second condition for the dominance, we do not create the arcs between the jobs belonging to the front in the evaluated node (job 4 and job 5). We observe that a max flow of $m+p$ units is necessary to ensure all job executions and to satisfy the resource constraints. In that case, $\sigma'(p)$ is a feasible schedule and we can prune the node.
\FloatBarrier
We propose a second dominance rule based on the position of the front jobs in the priority list. For a given schedule, the dominance rule searches for a new list of jobs in order to obtain the dominant partial schedule. We modify the list of scheduled jobs taking into account the precedence constraints. We can prune the evaluated node when the dominant partial schedule keeps the same front than the evaluated node (jobs 1, 2, and 3), one of the jobs starts earlier ($S'_{1}<S_{1}$) and for the rest of jobs belonging to the front the start times are not delayed ($S_{2}=S'_{2}$ and $S_{3}=S'_{3}$), as we see in Figure~\ref{fig-dominance}.
\FloatBarrier
We propose to permute the order of the $m$ front jobs in order to find the dominant schedule. For example, in Figure~\ref{fig-dominance} if the order of scheduled front jobs is $1-2-3$ we test all the possible permutations satisfying precedence constraints. If one of such permutations yields a dominant partial schedule, we can prune the evaluated node. This rule can be computed with time complexity $O(m!)$. As shown in Section~\ref{Evaluation}, despite its exponential worst-case complexity, this dominance rule has interesting properties when used in conjunction with discrepancy-based tree search and remains efficient for a small number of machines. A partial enumeration remains valid if $m$ becomes very large.
  
Note similar dominance rules have already been used for the RCPSP (which can be defined as an extension of the parallel machine scheduling problem with precedence constraints, but without setup times) under the name "cutset dominance rules"~\citep{bib-DominanceRules}. However, in~\cite{bib-DominanceRules}, all the cutsets are kept in memory yielding important memory requirements.
\section{Discrepancy-based tree search methods}
\subsection{Limited discrepancy search}\label{Section-LDS}
~~~~To tackle the combinatorial explosion of the standard branch-and-bound methods for large problem instances, we use a method based on the discrepancies regarding a reference branching heuristic. Such a method is based on the assumed good performance of this reference heuristic, thus making an ordered local search around the solution given by the heuristic. First, it explores the solutions with few discrepancies from the heuristic solution and then it moves away from this solution until it has covered the whole search space. In this context, the principle of \emph{LDS (Limited Discrepancy Search)}~\citep{bib-lds} is to explore first the solutions with discrepancies on top of the tree, since it assumes that the early mistakes, where very few decisions have been taken, are the most important. 

\begin {figure}[h]
\begin{center}
\includegraphics[width=13cm,height=6cm]{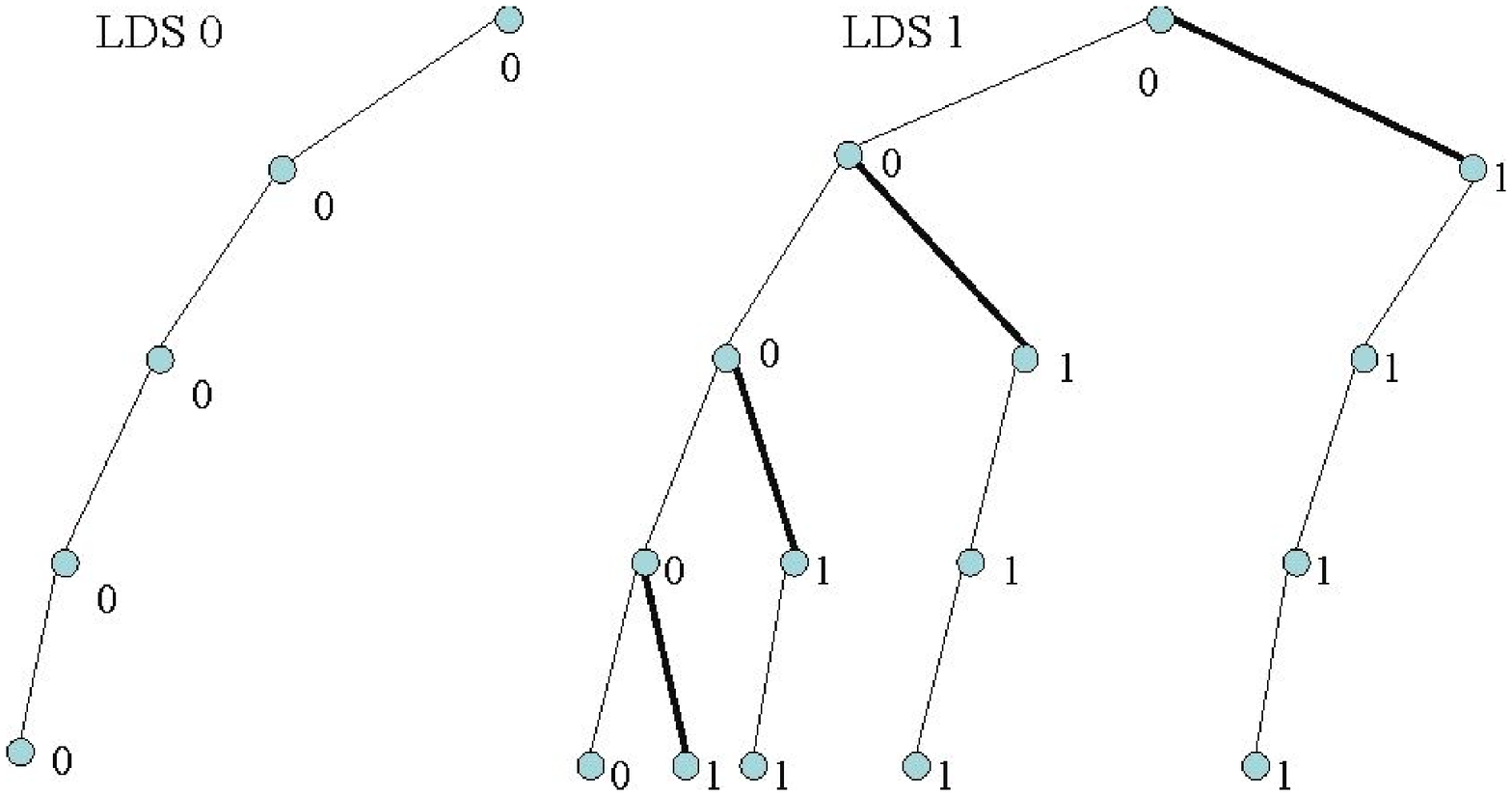}
\caption{\emph{Limited Discrepancy Search} for a binary tree}
\label{fig-LDS}
\end{center}
\end {figure}

Figure~\ref{fig-LDS} shows \emph{LDS} behavior for a binary tree search with the number of discrepancies for every node. Let us consider the left branch as the reference heuristic decision. At iteration 0 we explore the heuristic solution, then at iteration 1 we explore all the solutions that differ at most once from the heuristic solution, and we continue until all the leaves have been explored. 

LDS can be used as an exact method, for small-size instances, when the maximum number of discrepancies is authorized. We can also use it as an approximate method if we limit the number of authorized discrepancies.

Several methods based on LDS have been proposed to improve its efficiency. \emph{ILDS (Improved LDS)}~\citep{bib-ilds} has been devised to avoid the redundancy (observed in Figure~\ref{fig-LDS}) where the solutions with no discrepancies are also visited at iteration 1. \emph{DDS (Depth-bounded Discrepancy Search)}~\citep{bib-dbds} or \emph{DBDFS (Discrepancy-Bounded Depth First Search)}~\citep{bib-dbdfs} propose to change the order of the search. DDS limits the depth where the discrepancies are considered, in the sense that at the $k^{\rm th}$ iteration we only authorize the discrepancies at the first $k$ levels of the tree. It stresses the principle that the early mistakes are the most important. DBDFS consists in a classical \emph{DFS} where the nodes explored are limited by the discrepancies. Recently, in the \emph{YIELDS} method~\citep{bib-Yields}, learning process notions are integrated. In what follows, we propose several versions of LDS adapted to the considered parallel machine scheduling context.
\FloatBarrier
\subsection{Exploration strategy}
~~~~As a branching heuristic, we use the same heuristic to compute the lower bound presented in Section~\ref{subsec-evaluation}:~\emph{EST (Earliest Start Time)} rule for the selection of the next job to schedule and the resource to execute it. We take criterion EST because it is intuitively compatible with the minimization of setup times which has globally a positive impact for minimization of other regular criteria~\citep{bib-EST}. In case of tie between two jobs, we apply \emph{SPT (Smallest Processing Time)} rule for $\min \sum C_{i}$ and \emph{EDD (Earliest Due Date)} for $\min L_{\max}$.

\begin {figure}[h]
\begin{center}
\subfigure[binary]{
\includegraphics[scale=0.25]{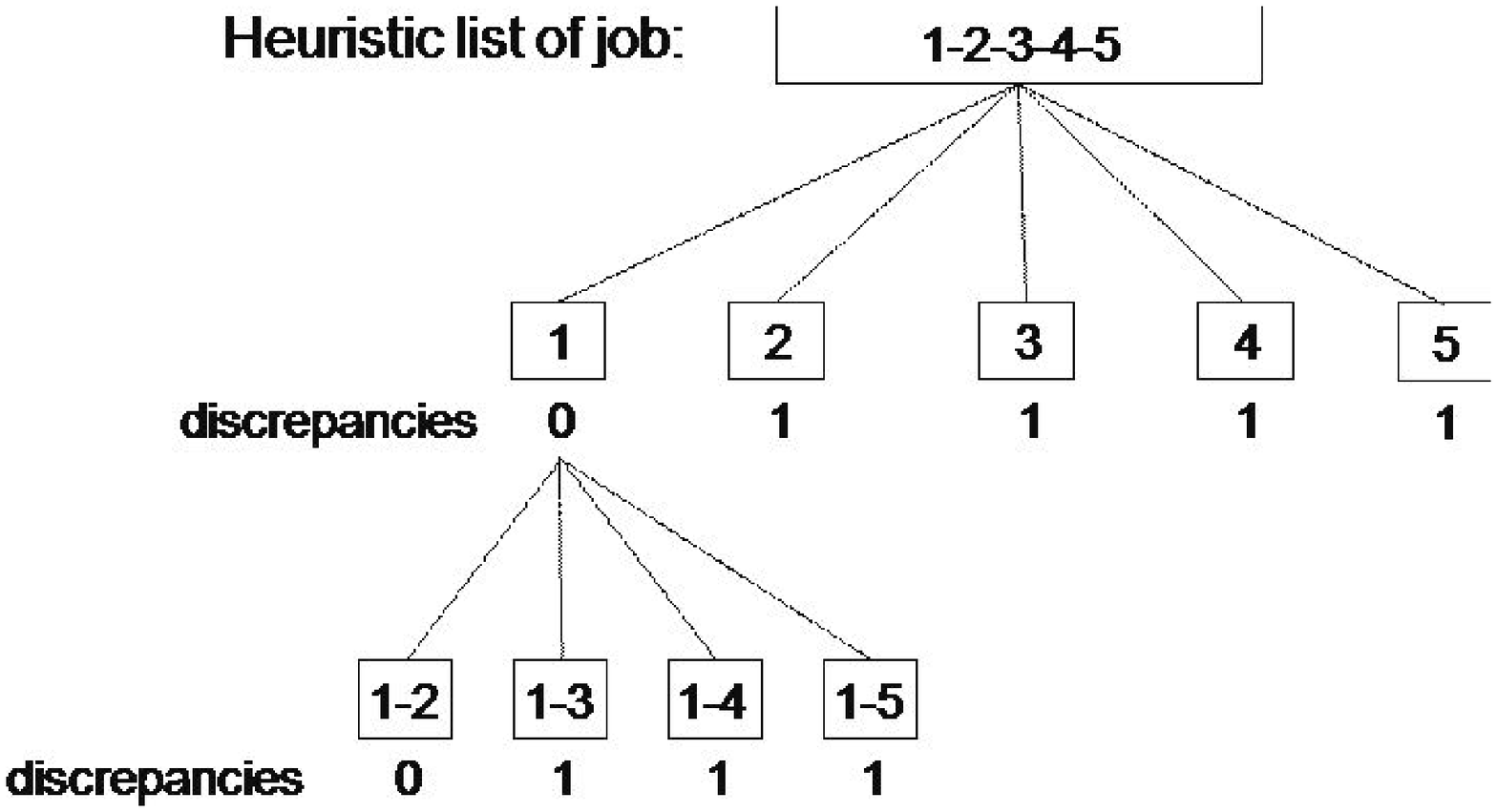}
%\caption{binary}
}
\subfigure[non-binary]{
\includegraphics[scale=0.25]{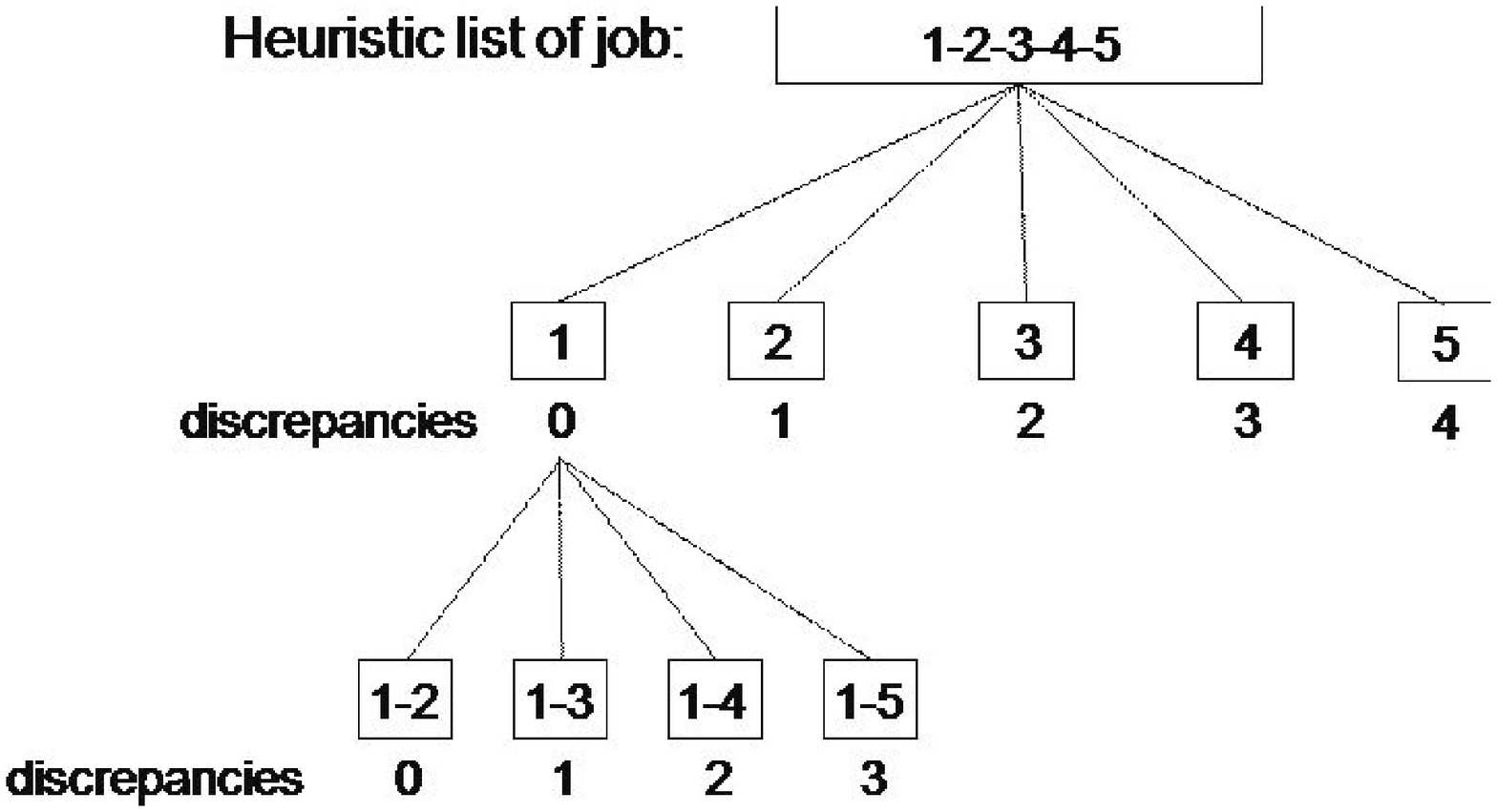}
%\caption{Non-binary counting mode}
}
\caption{Example of discrepancies counting modes on job selection}
\label{fig-modes}
\end{center}
\end {figure}

\begin {figure}[h]
\begin {center}
\scalebox{0.4}{
\includegraphics[]{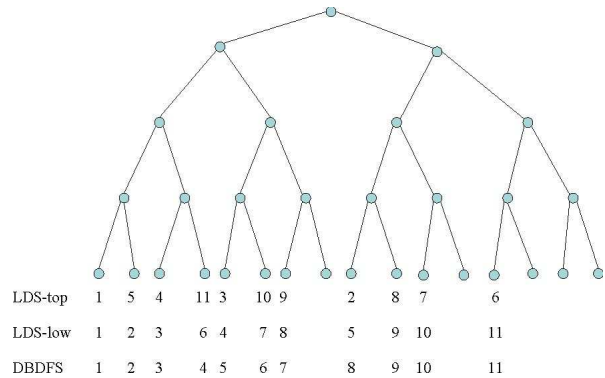}}
\caption{Order of explored leaves for different branching rules}
\label{fig-OrdreRecherche}
\end {center}
\end {figure}

Because of the existence of two types of decisions, we consider here two types of discrepancies: discrepancy on job selection and discrepancy on resource allocation. In the case of non-binary search trees, we have two different ways to count the discrepancies (see Figure~\ref{fig-modes}). In the first mode (\emph{binary}), we consider that choosing the heuristic decision corresponds to 0 discrepancy, while any other value corresponds to 1 discrepancy. The other mode (\emph{non-binary}) consists in considering that the further we are from the heuristic choice the more discrepancies we have to count. We suggest to evaluate experimentally both modes for the heuristic for job selection. On the other hand, for the choice of the machine, we use the non-binary mode since we assume that the allocation heuristic only makes a few errors. As we will see in Section~\ref{Evaluation}, selecting the machine which allows the earliest completion of the job is a high performance heuristic.
\FloatBarrier
We propose to test three different branching schemes. The first one, called DBDFS~\citep{bib-dbdfs}, is a classical depth-first search where the solutions obtained are limited by the allowed discrepancies (see Section~\ref{Section-LDS}). We propose two other strategies, \emph{LDS-top} and \emph{LDS-low}, which consider the number of discrepancies for the order in which the solutions are reached. The node to explore is the node with the smallest number of discrepancies, and with the smallest depth for the strategy called LDS-top, and with the largest depth for the strategy called LDS-low. As Figure~\ref{fig-OrdreRecherche} shows (case of $2$ authorized discrepancies) all three methods explore the same solutions but in different orders.
\FloatBarrier
%~\\[-5mm]
\subsection{Large neighborhood search based on LDS}   
We have presented \emph{LDS} as an exact or a truncated tree search method. In this section, we propose to use it as part of local search. In a local search method, we define a solution neighborhood $N_{k} (x)$ ($k$ defines the acceptable variations of solution $x$). If we find a solution $x'$ better than $x$ in $N_{k}(x)$ then we explore the neighborhood $N_k(x')$ of this new best solution. In the case of large-scale neighborhoods problems, the neighborhood becomes so huge that we can consider the search for the best solution in  $N_{k} (x)$ as an optimization sub-problem~\citep{bib-LNS}. In that context, we consider a neighborhood defined by an LDS search tree.
 
\emph{CDS (Climbing Discrepancy Search)}~\citep{bib-cds} is the first large neighborhood search method based on \emph{LDS} (see Algorithm~\ref{algo:CDS}). At each iteration it carries out a $k$-discrepancy search around the best current solution. If a better solution is found, then \emph{CDS} explores its neighborhood. In the case of no better solution is found, then $k$ is increased by one.
\begin{algorithm}[H]
\SetVline
\Begin{
  $k \leftarrow 1$\; 
  $k_{max}\leftarrow n$\;
  $Sol_{ref} \leftarrow InitialHeuristic()$\;
  \While {$k\leq k_{max}$}{
  \texttt{\footnotesize/* Generate the set of solutions $N$ of $k$ discrepancies from $Sol_{ref}$~*/}\\
  $N=LDS(Sol_{ref},k)$\;  	
  $s' \leftarrow BestOf(N)$\;		  
 \eIf{$z(s')<z(Sol_{ref})$}{
  $Sol_{ref}\leftarrow s'$\; 	
  $k \leftarrow 1$\;
  	       }{
$k \leftarrow k+1$\;
  		   }
	} 
}
\caption{\emph{Climbing Discrepancy Search}}
\label{algo:CDS}
\end{algorithm}
\FloatBarrier
The drawback of CDS is that for large-size instances the neighborhood quickly explodes.~\citet{bib-cdds} propose \emph{CDDS (Climbing Depth-bounded Discrepancy Search)} that mixes principles of CDS and of DDS. The neighborhood of the best solution is limited not only by the number of discrepancies but also by the depth in the tree. In that case, the neighborhood explosion is avoided and the idea that the most important heuristic mistakes are early ones is stressed.
 
In this work, we propose two variants of CDS and CDDS for the problems at hand. They are closely related with \emph{VNS (Variable Neighborhood Search)}~\citep{bib-VNS} concept, since we modify the size and the structure of the neighborhood explored.~\emph{HD-CDDS (Hybrid Discrepancy CDDS)} (see Algorithm~\ref{algo:HDCDDS}) consists in a mix of CDS and CDDS. We start with a CDS search, but if for a defined number of discrepancies $k_{limit}$ we cannot find a better solution, then we authorize a bigger number of discrepancies only between some levels ([$d_{min}$,$d_{max}$]). Once we have finished the search for $k_{limit}+1$, we propose either to increase the number of authorized discrepancies and to keep the same number of levels where the discrepancies are authorized ($x=d_{max}-d_{min}$), which is the case in Algorithm~\ref{algo:HDCDDS}, or to increase the number of levels and to keep the number of discrepancies. This method solves the problem of neighborhood explosion and offers more jobs mobility than CDDS (which is particularly interesting for setup times problems) but we need to parametrize the values of the search ($k_{limit}$, $x$). 
\begin {algorithm}[H]
\SetVline
\Begin{
  $k \leftarrow 1$\;
  $d_{min}\leftarrow 0$\;
  $d_{max}\leftarrow n$\; 
  $Sol_{ref} \leftarrow InitialHeuristic()$\;
  \While{termination conditions not met}{
  \texttt{\footnotesize/* Generate the set of solutions $N$ of $k$ discrepancies from $Sol_{ref}$~*/}\\
  $N=GenSol(Sol_{ref},k,d_{min},d_{max})$\;
  $s' \leftarrow BestOf(N)$\;		  
  \eIf{$z(s')<z(Sol_{ref})$}{
  $Sol_{ref}\leftarrow s'$\; 	
  $k \leftarrow 1$\;
  $d_{min}\leftarrow 0$\;
  $d_{max}\leftarrow n$\;
  }{
  \eIf{$k<k_{limit}$}{
   $k \leftarrow k+1$\;
   }{
  \eIf{$d_{max}-d_{min}=n$}{
  $d_{min}\leftarrow 0$\;
  $d_{max}\leftarrow x$\;
  }{
  $d_{min}\leftarrow d_{max}$\;
  $d_{max}\leftarrow d_{min}+x$\;
  \If{$d_{min}>n$}{
  $k\leftarrow k+1$\;
  $d_{min}\leftarrow 0$\;
  $d_{max}\leftarrow x$\;}
  }
  }
  }
  }
} 
\caption{Algorithm \emph{HD-CDDS}}
\label{algo:HDCDDS}
\end{algorithm}
\FloatBarrier
The second proposed variant, \emph{MC-CDS (Mix Counting CDS)}, is an application of \emph{CDS} but with a modification in the way to count the discrepancies for the job selection rule only. We consider a binary counting for the discrepancies at the top level of the tree and a non-binary counting way for the rest of levels. This variant accepts discrepancies for all depth levels because the non-binary counting restricts the explored neighborhood.
\subsection{Discrepancy-adapted dominance rules}
~~~In this section we propose to adapt the second dominance rule presented in Section~\ref{subsec-dominance} to the principle of local search. We argue that it can be very inefficient to use the dominance rule as presented in Section~\ref{subsec-dominance} with the proposed local search methods. Indeed, the best solutions of the neighborhood could not be explored because we have found a dominant partial schedule that allows us to prune them. Even if it is true that there exists a solution better than the evaluated node, it may not belong to the explored neighborhood.
 
For that reason, we propose discrepancy-adapted dominance rules. Once we know the criterion that defines the neighborhood (for example, $k$ authorized discrepancies from the job list $L$), we only have to verify that the new list of jobs $L'$ that reaches the dominant partial schedule is part of the explored nodes in the local search ($L'\in G$, where $G$ is the set of $k$-discrepancies lists from $L$). 

We can see that the max flow computation rule presented in Section~\ref{subsec-dominance} is not discrepancy adaptable. It is not possible to verify that the dominant partial schedule $\sigma'(p)$ is part of the explored space because the rule indicates the existence of $\sigma'(p)$ but not the corresponding schedule. On the other hand, the second dominance rule introduced in Section~\ref{subsec-dominance} consists in a local modification of the evaluated schedule in order to explicitly obtain the dominant schedule. That way, we have the list of jobs, $L'$, available to compare with the best current solution list of jobs, $L$, and to verify that the dominant schedule is part of the explored nodes. Hence, when the encountered dominant schedule is not part of the explored neighborhood the current node is not pruned.              
\section{Computational experiments}\label{Evaluation}
~~~In this section we present the main results obtained from the implementation of our work. In the literature we have not found instances for parallel machines including both setup times and precedence constraints. Therefore, we propose to test the methods on a set of randomly generated instances. The algorithms are implemented in C++ and were run on a 2 GHz personal computer with 2 Go of RAM under the Linux Fedora 8 operating system.  

We generate a set of 120 (60 for each criterion) small-size instances ($n=10$, $m=3$, and $n=15$, $m=2$) for the evaluation of the dominance rules and for the \emph{ECT rule} efficiency. Then, we test on a set of 120 middle-size instances ($n=40$, $m\in[2,4]$) the different branching rules (\emph{LDS-top}, \emph{LDS-low}, and \emph{DBDFS}), the different ways to count the discrepancies (\emph{binary} and \emph{non-binary}) to determine the best methods for being included inside the \emph{LDS} structure of the local search methods. The efficiency of the lower bounds, the dominance rules and the energetic reasoning proposed in Section~\ref{sec-branchandbound} are tested on middle and large-size instances ($n=100$, $m\in[2,4]$). We also compare the \emph{CDS} and the \emph{HD-CDDS} methods with the results obtained in \cite{bib-TreeSearch} for the hard instances of the $Pm|r_{i},q_{i}|C_{\max}$ problem (without precedence constraints and setup times). And finally, we evaluate and compare the proposed methods on a set of 120 large-size instances with the results obtained with ILOG OPL 6.0.

We use the RanGen software~\citep{bib-rangen} in order to generate the precedence graph between the jobs. Setup times and time windows $[r_i, d_i]$ cannot be generated by RanGen. Setup times are generated from the uniform distributions $U[1,10]$ and $U[20,40]$
. Moreover they  must respect the weak triangle inequality: $s_{ij}\leq s_{ik}+p_{k}+s_{kj},\forall \thinspace i,j,k$. The values of $p_i$ are generated from the uniform distribution $U[1,5]$ %(when $s_{ij}\sim U[1,20]$) and $U[1,10]$ (when $s_{ij}\sim U[1,10]$)
. Time windows are generated in a classical way we found in the literature~\citep{bib-Sourd}. The values of $d_{i}$ are generated from the uniform distribution $U[\max(0,P\times(1-\tau-\rho/2)),P\times(1-\tau+\rho/2)]$, where $P=\sum(p_{i}+\min_{j}(s_{ij}))$, $\tau\in[0,1]$, $\rho\in[0,1]$. The $r_{i}$ are generated from $d_i$, $r_{i}=d_{i}-(p_{i}\times(2+\alpha))$ where $\alpha \in [-0.5,+1.5]$.

We solve to optimality the small-size instances and we compare the results (\emph{Optimal}) with the results obtained when we apply the ECT rule (\emph{ECT}) for each possible list of jobs (jobs are only allocated to the machine which allows to finish it first), with the results using the dominance rule based on the permutation of front jobs (\emph{Front Rule}), and with the results using the dominance rule based on max flow computation (\emph{Max Flow}).             
\begin {table}[hbt]
\begin{center}
\subtable{
\begin {tabular}{lccc}
%\hline
\multicolumn{1}{l}{60 Instances}& & &\\
\multicolumn{1}{l}{\footnotesize{$n=10,m=3$}} 
&\emph{NbBest} & \emph{AvgNodes} & \emph{AvgTCPU}\\
\hline
\emph{Optimal} & 60 (100.0~\textdiscount) & 484925 & 10.6 \\
%\hline
\emph{Front Rule} & 60 (100.0~\textdiscount)  & 480444 & 12.3\\
%\hline
\emph{Max Flow} & 60 (100.0~\textdiscount) &  339541    &  27.7     \\
%\hline
\emph{ECT} & 53 (88.3~\textdiscount)	& 61684 & 0.07 \\
%\hline
\end {tabular}
}
\subtable{
\begin {tabular}{lccc}
%\hline
\multicolumn{1}{l}{60 Instances}& & &\\
\multicolumn{1}{l}{\footnotesize{$n=15,m=2$}} 
&\emph{NbBest} & \emph{AvgNodes} & \emph{AvgTCPU}\\
\hline
\emph{Optimal} & 60 (100.0~\textdiscount) & 10126793 & 641.9 \\
%\hline
\emph{Front Rule} & 60 (100.0~\textdiscount)  & 9480313  & 626.4\\
%\hline
\emph{Max Flow} & 60 (100.0~\textdiscount) &   7530154   & 454.6     \\
%\hline
\emph{ECT} & 54 (90.0~\textdiscount)	& 1747416 & 2.5 \\
%\hline
\end {tabular}
}
%~\\[3mm]
\caption{Results of ECT and dominance rules efficiency for $\min \sum{C_i}$ problem}
\label{fig-TestECT}
\end {center}
\end {table}
\begin {table}[hbt]
\begin{center}
\subtable{
\begin {tabular}{lccc}
%\hline
\multicolumn{1}{l}{60 Instances}& & & \\ 
\multicolumn{1}{l}{\footnotesize{$n=10,m=3$}} 
&\emph{NbBest} & \emph{AvgNodes} & \emph{AvgTCPU}\\
\hline
\emph{Optimal} & 60 (100.0~\textdiscount) & 281896 & 5.6 \\
%\hline
\emph{Front Rule} & 60 (100.0~\textdiscount)  & 263474 & 7.9\\
%\hline
\emph{Max Flow} & 60 (100.0~\textdiscount) & 219557    & 19.7       \\
%\hline
\emph{ECT} & 52 (86.7~\textdiscount)	& 69141 & 0.07\\
%\hline
\end {tabular}
}
\subtable{
\begin {tabular}{lccc}
%\hline
\multicolumn{1}{l}{60 Instances}& & &\\
\multicolumn{1}{l}{\footnotesize{$n=15,m=2$}} 
&\emph{NbBest} & \emph{AvgNodes} & \emph{AvgTCPU}\\
\hline
\emph{Optimal} & 60 (100.0~\textdiscount) & 11936385  & 884.8 \\
%\hline
\emph{Front Rule} & 60 (100.0~\textdiscount)  & 10503767 & 778.7\\
%\hline
\emph{Max Flow} & 60 (100.0~\textdiscount) &    8945948  & 628.4       \\
%\hline
\emph{ECT} & 54 (90.0~\textdiscount)	& 4681104 & 7.27 \\
%\hline
\end {tabular}
}
%~\\[3mm]
\caption{Results of ECT and dominance rules efficiency for $\min{L_{\max}}$ problem}
\label{fig-TestECTb}
\end {center}
\end {table}
\FloatBarrier 
First, note that we found some hard instances that we could not to solve to optimality before 15000 seconds.  We observe in Tables~\ref{fig-TestECT} and~\ref{fig-TestECTb} that ECT rule is very efficient for both problems. The optimal solution is reached over almost 90~\textdiscount~of the instances and the average CPU time (\emph{AvgTCPU}) is clearly reduced when we use the ECT rule. These results let us consider, for local search methods, only the job permutation allocating the jobs on the machines following the ECT rule. The dominance front rule is also effective, the average number of explored nodes (\emph{AvgNodes}) and the average CPU time usually decrease when we use it. We observe that the \emph{Max Flow} rule largely reduces the number of explored nodes and the CPU time, except for the very small-size instances. We deduce that it is a very efficient rule to solve to optimality instances with a larger number of jobs.     

In the comparison between the two different ways to count the discrepancies, \emph{binary} and \emph{non-binary} (only for job selection rules), we have evaluated on the middle-size instances the number of times each mode has found the best solution (\emph{NbBest}). The CPU time is limited to $100$ seconds.

Table~\ref{fig-TestModes} shows that the binary mode has a higher performance than the non-binary one. Out of a set of 120 instances, the binary mode has found the best solution over 75~\textdiscount~of the instances, independently of the branching rule. We find very similar results for both criteria. In the following, the binary counting is kept for the LDS structure of the local search. 
\begin {table}[hbt]
\begin{center}
\begin {tabular}{lcc}
%\hline
\multicolumn{1}{l}{120 Instances}& 
\multicolumn{2}{c}{\emph{NbBest}}\\
\multicolumn{1}{l}{\footnotesize{$n=40,m\in[2,4]$}}&
\emph{binary} mode & \emph{non-binary} mode\\
\hline
\emph{DBDFS} & 90 (75.0~\textdiscount) & 48 (40.0~\textdiscount)\\
%\hline
\emph{LDS-top} & 93 (77.5~\textdiscount)	& 49 (40.8~\textdiscount)\\
%\hline
\emph{LDS-low} & 98 (81.7~\textdiscount)  & 31 (25.8~~\textdiscount)\\
%\hline
\end {tabular}
%~\\[3mm]
\caption{Results of the comparison between discrepancies counting modes}
\label{fig-TestModes}
\end {center}
\end {table}  
\FloatBarrier   
In Table~\ref{fig-TestBranchingRule}, we can see the results for the comparison between the exploration strategies. In addition to previous notations, we introduce the average mean deviation from the best solution (\emph{AvgDev}). The CPU time is limited to $100$ seconds.  
\begin {table}[hbt]
\begin{center}
\begin {tabular}{lcccc}
%\hline
\multicolumn{1}{l}{Binary mode}& 
\multicolumn{2}{c}{$\min \sum{C_i}$~(60 instances)}&
\multicolumn{2}{c}{$\min{L_{\max}}$~(60 instances)}\\
\multicolumn{1}{c}{\footnotesize{$n=40,m\in[2,4]$}}& \multicolumn{1}{c}{\emph{NbBest}} &\multicolumn{1}{c}{\emph{AvgDev}}& \multicolumn{1}{c}{\emph{NbBest}} &\multicolumn{1}{c}{\emph{AvgDev}}\\
\hline
\emph{DBDFS} & 43 (71.7~\textdiscount) & 0.91~\textdiscount& 47 (78.3~\textdiscount) & 1.86~\textdiscount  \\
%\hline
\emph{LDS-top} & 29 (48.3~\textdiscount) & 0.43~\textdiscount& 17 (28.3~\textdiscount) & 2.33~\textdiscount  \\
%\hline
\emph{LDS-low} & 50 (83.3~\textdiscount) & 0.71~\textdiscount& 59 (98.3~\textdiscount) & 0.75~\textdiscount  \\
%\hline
\end {tabular}
\caption{Results for the comparison of different branching strategies}
\label{fig-TestBranchingRule}
\end {center}
\end {table}
\FloatBarrier
We find that LDS-low is the most efficient strategy, since it reaches the best solution for a larger number of instances and it presents the less important average mean deviation when the best solution is found by another strategy. LDS-low finds the best solution for all instances except for one corresponding to the maximum lateness minimization and for 50 over a set of 60 instances for completion times sum minimization. We use this strategy for the remaining computational experiments.

The lower bounds, the energetic reasoning, and the discrepancy-adapted dominance rule are compared in Tables~\ref{fig-Regles} and ~\ref{fig-Reglesb}. We run a 30 seconds LDS search for the middle and large-size instances for different versions of the node evaluation. First, we only consider the lower bound computed using precedence constraint propagation (\emph{LBCP}), then we add the lower bound ($LB_{NCY}$) proposed in~\cite{bib-lbCi} for $\min \sum{C_i}$ problem and the energetic reasoning (\emph{ENERGY}) for $\min L_{\max}$ problem; finally we add the discrepancy-adapted dominance rule (\emph{DaDR}). We compare the number of times each version finds the best solution (\emph{NbBest}), the explored nodes average (\emph{AvgNodes}), and the average CPU time needed to reach the best solution (\emph{TBest}), only for the cases that all versions have found it.
\begin {table}[hbt]
\begin{center}
\subtable{
\begin {tabular}{lccc}
%\hline
\multicolumn{1}{l}{60 Instances}& & &\\
\multicolumn{1}{l}{\footnotesize{$n=40,m\in[2,4]$}} 
&\emph{NbBest} & \emph{AvgNodes} & \emph{TBest}\\
\hline
\emph{LBCP} & 36 (60.0~\textdiscount) & 62007 &4.52 \\
%\hline
$LB_{NCY}$ & 38  (63.3~\textdiscount)  & 61742 &4.47\\
%\hline
\emph{DaDR} & 35 (58.3~\textdiscount)	&  53373  &1.69\\
%\hline
\end {tabular}
}
\subtable{
\begin {tabular}{lccc}
%\hline
\multicolumn{1}{l}{60 Instances}& & &\\
\multicolumn{1}{l}{\footnotesize{$n=100,m\in[2,4]$}} 
&\emph{NbBest} & \emph{AvgNodes}& \emph{TBest}\\
\hline
\emph{LBCP} & 26 (43.3~\textdiscount) & 9259 & 17.55 \\
%\hline
$LB_{NCY}$ & 34 (56.7~\textdiscount)  & 7813 &15.63\\
%\hline
\emph{DaDR} & 38 (63.3~\textdiscount)	& 7606 & 8.71\\
%\hline
\end {tabular}
}%~\\[3mm]
\caption{Results of lower bounds and dominance rule efficiency for $\min \sum{C_i}$ problem}
\label{fig-Regles}
\end {center}
\end {table}
\FloatBarrier

Tables \ref{fig-Regles} and \ref{fig-Reglesb} show the efficiency of the specific lower bound $LB_{LCY}$ and energetic reasoning with the computation of setup times consumption. Moreover, we find that the discrepancy-adapted dominance rule is very efficient for large-size instances but not especially interesting for the middle-size instances. However the time consumed to reach the best solution is reduced when we use the dominance rule for most of cases.
\begin {table}[hbt]
\begin{center}
\subtable{
\begin {tabular}{lccc}
%\hline
\multicolumn{1}{l}{60 Instances}& & &\\
\multicolumn{1}{l}{\footnotesize{$n=40,m\in[2,4]$}} 
&\emph{NbBest} & \emph{AvgNodes}& \emph{TBest}\\
\hline
\emph{LBCP} & 47 (78.3~\textdiscount) & 93737& 4.81  \\
%\hline
\emph{ENERGY} & 48 (80.0~\textdiscount)  & 99856& 4.24 \\
%\hline
\emph{DaDR} & 44 (73.3~\textdiscount)	& 71737  & 4.59 \\
%\hline
\end {tabular}
}
\subtable{
\begin {tabular}{lccc}
%\hline
\multicolumn{1}{l}{60 Instances}& & &\\
\multicolumn{1}{l}{\footnotesize{$n=100,m\in[2,4]$}} 
&\emph{NbBest} & \emph{AvgNodes}& \emph{TBest}\\
\hline
\emph{LBCP} & 44 (73.3~\textdiscount) & 11474& 4.29  \\
%\hline
\emph{ENERGY} & 48 (80.0~\textdiscount)  & 12961& 3.58 \\
%\hline
\emph{DaDR} & 55 (91.7~\textdiscount) & 9462& 3.17 \\
%\hline
\end {tabular}}
%~\\[3mm]
\caption{Results of lower bound, energetic reasoning and dominance rule efficiency for $\min{L_{\max}}$ problem}
\label{fig-Reglesb}
\end {center}
\end {table}    
\FloatBarrier
We compare \emph{CDS} and \emph{HD-CDDS} methods against other tree search methods presented in~\cite{bib-TreeSearch}. In~\cite{bib-TreeSearch}, the authors test two different branching schemes, time windows (\emph{tw}) and chronological (\emph{chr}), and several incomplete tree search techniques (truncated branch-and-bound, LDS, Beam Search and Branch-and-Greed) for the $Pm|r_{i},q_{i}|C_{\max}$ problem. We adapt the proposed methods for this problem and we use the heuristic for the initial solution and the upper bounds proposed in their paper. In Table~\ref{fig-ComparaisonNeron}, we compare LDS ($z$ is the number of authorized discrepancies) and Beam Search (\emph{BS}, $\omega$ is the number of explored child nodes) results, the method with the best results in their work, against the proposed methods CDS and HD-CDDS. We have evaluated the number of times the method has found the best solution (\emph{NbBest}) and for how many of them the method is the only one to reach the best solution (\emph{NbBestStrict}) for a set of 50 hard instances ($n=100$ and $m=10$). The CPU time is limited to $30$ seconds as in~\cite{bib-TreeSearch}.
\begin {table}[hbt]
\begin{center}
\begin{tabular}{l c c}
    %\hline
    50 instances  & \emph{NbBest} & \emph{NbBestStrict}\\
    \hline
    \small{$LDS^{tw}_{z=1}$} & 1 (2.0~\textdiscount)  & 0 \\
    \small{$LDS^{chr}_{z=2}$} & 7 (14.0~\textdiscount)  & 0 \\
    \small{$BS^{tw}_{\omega =3}$}&  25 (50.0~\textdiscount)  & 3 \\
    \small{$BS^{chr}_{\omega=4}$} & 22 (44.0~\textdiscount)  & 0\\
    \hline
    \emph{CDS} & 35 (70.0~\textdiscount)  &  6 \\
    \emph{HD-CDDS}&  38 (76.0~\textdiscount)  & 9\\    
    %\hline
\end{tabular}
\caption{Results for the comparison with other truncated tree search techniques}
\label{fig-ComparaisonNeron}
\end{center}
\end{table}
\FloatBarrier   
Although precedence constraints and setup times are not considered in the problem, we can observe that our propositions are strictly better. Out of a set of 50 instances, CDS and HD-CDDS find the best solution for most of the cases and they find a new best solution for 6 and 9 instances respectively. Rather than contradicting the statement of relative LDS inefficiency for parallel machine problem experienced by~\cite{bib-TreeSearch}, this demonstrates, at least for this problem, the efficiency of large neighborhood search based on LDS. 

Finally, we compare the local search methods with the results obtained by ILOG OPL 6.0. The four variants of the hybrid tree local search methods (\emph{CDS}, \emph{CDDS}, \emph{HD-CDDS}, \emph{MC-CDS}) are implemented with \emph{LDS-low}, discrepancy-adapted dominance rule and binary counting (except for \emph{MC-CDS} which supposes a mix counting). We solve the large-size instances ($n=100,m\in[2,4]$) for two different CPU time limits, 30 and 300 seconds, then we compare the number of times when the best solution has been found by the method and the average deviation from the best solution.
\begin {table}[hbt]
\begin{center}
\subtable{
\begin {tabular}{lcccc}
%\hline
\multicolumn{1}{l}{30 instances}& 
\multicolumn{2}{c}{\footnotesize{$TCPU=30s$}}&\multicolumn{2}{c}{\footnotesize{$TCPU=300s$}}\\
\multicolumn{1}{c}{\scriptsize{$p\sim U[1,5], s_{ij}\sim U[1,10]$}}& \multicolumn{1}{c}{\emph{NbBest}} &\multicolumn{1}{c}{\emph{AvgDev}} & \multicolumn{1}{c}{\emph{NbBest}} &\multicolumn{1}{c}{\emph{AvgDev}}\\
\hline
\emph{CDS} & 17 (56.6~\textdiscount) & 0.64~\textdiscount & 7 (23.3~\textdiscount) & 0.51~\textdiscount  \\
%\hline
\emph{CDDS} & 7 (23.3~\textdiscount) & 0.75~\textdiscount & 7 (23.3~\textdiscount) & 0.82~\textdiscount  \\
%\hline
\emph{HD-CDDS} & 16 (53.3~\textdiscount) & 0.60~\textdiscount & 14 (46.7~\textdiscount) & 0.43~\textdiscount  \\
%\hline
\emph{MC-CDS} & 17 (56.6~\textdiscount) & 0.64~\textdiscount & 10 (33.3~\textdiscount) & 0.45~\textdiscount  \\
%\hline
\emph{ILOG OPL} & 4 (13.3~\textdiscount) & 1.51~\textdiscount & 2 (6.7~\textdiscount) & 1.47~\textdiscount  \\
%\hline
\end {tabular}
}
\subtable{
\begin {tabular}{lcccc}
%\hline
\multicolumn{1}{l}{30 instances}& 
\multicolumn{2}{c}{\footnotesize{$TCPU=30s$}}&\multicolumn{2}{c}{\footnotesize{$TCPU=300s$}}\\
\multicolumn{1}{c}{\scriptsize{$p\sim U[1,5], s_{ij}\sim U[20,40]$}}& \multicolumn{1}{c}{\emph{NbBest}} &\multicolumn{1}{c}{\emph{AvgDev}} & \multicolumn{1}{c}{\emph{NbBest}} &\multicolumn{1}{c}{\emph{AvgDev}}\\
\hline
\emph{CDS} & 9 (30.0~\textdiscount) & 0.23~\textdiscount & 6  (20.0~\textdiscount) & 0.18~\textdiscount  \\
%\hline
\emph{CDDS} & 7 (23.3~\textdiscount) & 0.35~\textdiscount & 6 (20.0~\textdiscount) & 0.38~\textdiscount  \\
%\hline
\emph{HD-CDDS} & 12 (40.0~\textdiscount) & 0.26~\textdiscount & 11  (36.6~\textdiscount) & 0.17~\textdiscount  \\
%\hline
\emph{MC-CDS} & 11 (36.7~\textdiscount) & 0.25~\textdiscount & 13  (43.3~\textdiscount) & 0.26~\textdiscount  \\
%\hline
\emph{ILOG OPL} & 10 (33.3~\textdiscount) & 0.70~\textdiscount & 5 (16.6~\textdiscount) & 0.63~\textdiscount  \\
%\hline
\end {tabular}
}
\caption{Results for the comparison of different variants of hybrid tree local search methods for $\min \sum{C_i}$ problem}
\label{fig-TestLocalSearch}
\end {center}
\end {table}

\FloatBarrier
In Table~\ref{fig-TestLocalSearch}, we observe that hybrid local search methods improve the best solutions found by ILOG OPL. All methods, except CDDS, find the best solution for a large number of instances and the mean deviation from the best solution are less important than ILOG OPL solutions. We observe that computing an upper bound highly increases the efficiency of the truncated search.  
\begin {table}[hbt]
\begin{center}
\subtable{
\begin {tabular}{lcccc}
%\hline
\multicolumn{1}{l}{30 instances}& 
\multicolumn{2}{c}{\footnotesize{$TCPU=30s$}}&\multicolumn{2}{c}{\footnotesize{$TCPU=300s$}}\\
\multicolumn{1}{c}{\scriptsize{$p\sim U[1,5], s_{ij}\sim U[1,10]$}}& \multicolumn{1}{c}{\emph{NbBest}} &\multicolumn{1}{c}{\emph{AvgDev}} & \multicolumn{1}{c}{\emph{NbBest}} &\multicolumn{1}{c}{\emph{AvgDev}}\\
\hline
\emph{CDS} & 10 (33.3~\textdiscount) & 2.75~\textdiscount & 7 (23.3~\textdiscount) & 3.06~\textdiscount  \\
%\hline
\emph{CDDS} & 9 (30.0~\textdiscount) & 2.65~\textdiscount & 8 (26.7~\textdiscount) & 3.28~\textdiscount  \\
%\hline
\emph{HD-CDDS} & 13 (43.3~\textdiscount) & 1.92~\textdiscount & 10 (33.3~\textdiscount) & 2.56~\textdiscount  \\
%\hline
\emph{MC-CDS} & 13 (43.3~\textdiscount) & 1.75~\textdiscount & 11 (30.0~\textdiscount) & 2.29~\textdiscount  \\
%\hline
\emph{ILOG OPL} & 15 (50.0~\textdiscount) & 2.07~\textdiscount & 18 (60.0~\textdiscount) & 1.55~\textdiscount  \\
%\hline
\end {tabular}
}
\subtable{
\begin {tabular}{lcccc}
%\hline
\multicolumn{1}{l}{30 instances}& 
\multicolumn{2}{c}{\footnotesize{$TCPU=30s$}}&\multicolumn{2}{c}{\footnotesize{$TCPU=300s$}}\\
\multicolumn{1}{c}{\scriptsize{$p\sim U[1,5], s_{ij}\sim U[20,40]$}}& \multicolumn{1}{c}{\emph{NbBest}} &\multicolumn{1}{c}{\emph{AvgDev}} & \multicolumn{1}{c}{\emph{NbBest}} &\multicolumn{1}{c}{\emph{AvgDev}}\\
\hline
\emph{CDS} & 3 (10.0~\textdiscount) & 2.76~\textdiscount & 2 (6.0~\textdiscount) & 2.89~\textdiscount  \\
%\hline
\emph{CDDS} & 3 (10.0~\textdiscount) & 2.71~\textdiscount & 2 (6.0~\textdiscount) & 2.88~\textdiscount  \\
%\hline
\emph{HD-CDDS} & 13 (43.3~\textdiscount) & 2.12~\textdiscount & 7 (23.3~\textdiscount) & 1.55~\textdiscount  \\
%\hline
\emph{MC-CDS} & 12 (40.0~\textdiscount) & 2.08~\textdiscount & 8 (26.7~\textdiscount) & 1.83~\textdiscount  \\
%\hline
\emph{ILOG OPL} & 15 (50.0~\textdiscount) & 0.91~\textdiscount & 19 (63.3~\textdiscount) & 0.90~\textdiscount  \\
%\hline
\end {tabular}
}
\caption{Results for the comparison of different variants of hybrid tree local search methods for $\min L_{\max}$ problem}
\label{fig-TestLocalSearchb}
\end {center}
\end {table}
\FloatBarrier
Table~\ref{fig-TestLocalSearchb} shows the results for the minimization of maximum lateness. For this case, we observe ILOG OPL improves our results, but we can say that the proposed methods are still competitive, the mean deviation is acceptable and they found the best solution over 50~\textdiscount~and 37~\textdiscount~of instances, for 30 and 300 seconds respectively.  

\section{Conclusion}
~~~In this paper we have studied limited discrepancy-based search methods. We have compared and tested some of the existing options for different LDS components, such as discrepancy counting modes and branching structures, to solve the parallel machine scheduling problem with precedence constraints and setup times. 

New local search methods based on LDS have been proposed and compared with similar existing methods. The computational experiments show these methods are efficient to solve parallel machine scheduling problems in general and demonstrates the interest, at least for the studied problem, of incorporating LDS into a large neighborhood search scheme as first suggested by~\cite{bib-cds}.

We have suggested an energetic reasoning scheme integrating setup times and we have proposed new global and local dominance rules adapted to discrepancies. As the results show, these evaluation techniques allow to reduce the number of explored nodes and the time of the search.

As a direction for further research, the proposed methods could be extended to solve more complex problems involving setup times, like the hybrid flow shop or the RCPSP.

\bibliographystyle{plainnat} 
\bibliography{bibliografia}
\end{document}